\documentclass[aps,prl,twocolumn,superscriptaddress,nofootinbib]{revtex4-2}

\usepackage{amsmath,amssymb,amsfonts,mathtools}
\usepackage{physics}
\usepackage{braket}
\usepackage{bm}
\usepackage{mathrsfs}
\usepackage{tikz-cd}
\usepackage{graphicx}
\usepackage{subcaption}  
\newcommand{\prlsection}[1]{{\bf{#1}.---~}}
\newcommand{\DiamondNorm}[1]{\left\lVert #1\right\rVert_{\diamond}}
\usepackage[
  breaklinks    = true,
  colorlinks    = true,
  hypertexnames = false,
  pdfpagelabels = false,
  citecolor     = {blue!80!black},
  linkcolor     = {blue!80!black},
  urlcolor      = {blue!80!black},
]{hyperref} 
\usepackage[capitalize,nameinlink]{cleveref} 

\begin{document}

\title{Mereological Quantum Phase Transitions}

\author{Paolo Zanardi}
\affiliation{
Department of Physics and Astronomy, and Center for Quantum Information Science and Technology, University of Southern California, Los Angeles, California 90089-0484, USA}

\affiliation{Department of Mathematics, University of Southern California, Los Angeles, California 90089-2532, USA}
\author{Emanuel Dallas}
\affiliation{
Department of Physics and Astronomy, and Center for Quantum Information Science and Technology, University of Southern California, Los Angeles, California 90089-0484, USA}

\author{Faidon Andreadakis}
\affiliation{
Department of Physics and Astronomy, and Center for Quantum Information Science and Technology, University of Southern California, Los Angeles, California 90089-0484, USA}

\date{\today}

\begin{abstract}
We introduce the novel concept of {\em mereological quantum phase transitions} (m-QPTs). Our framework is based on a variational family of operator algebras defining generalized tensor product structures (g-TPS), a parameter-dependent Hamiltonian, and a quantum scrambling functional. By minimizing the scrambling functional, one selects a g-TPS, enabling a pullback of the natural information-geometric metric on the g-TPS manifold to the parameter space. The singularities of this induced metric -- so-called algebra susceptibility -- in the thermodynamic limit characterize the m-QPTs. We illustrate this framework through analytical examples involving quantum coherence and operator entanglement. Moreover, spin-chains numerical simulations  show susceptibility sharp responses at an integrability point and strong growth across disorder-induced localization, suggesting critical reorganizations of emergent subsystem structure aligned  with those transitions.
\end{abstract}

\maketitle
\prlsection{Introduction}
Mereology is a field of conceptual inquiry which dates back to Plato and Aristotle. It is concerned with 
the formal theory of the parthood relation, between parts and wholes, typically via axiomatic systems \cite{varziMereology2019}; this includes part–whole, part–part within a whole, and associated notions like overlap and fusion.
Quantum theory, with the associated conceptual puzzles of non-separability and quantum entanglement, imbues mereological investigation with extra layers of subtlety and complexity, see, e.g., \cite{Esfeld2004Relations,CalosiFanoTarozzi2011,Caulton2015Fermions,Nager2021MereologicalProblem,Nager2025Taxonomy}.
Recently, there has been growing interest in {\em{quantum mereology}} based on  quantum information and  operational ideas \cite{carrollQuantumMereologyFactorizing2021,zanardiOperationalQuantumMereology2024,LoizeauSels2025QMereologySpectrum,Soulas2025DisentanglingTPS}.

In particular, Ref. \cite{zanardiOperationalQuantumMereology2024} explored an approach based on operator algebras and quantum scrambling \cite{HaydenPreskill2007BlackHolesAsMirrors,SekinoSusskind2008FastScramblers,ShenkerStanford2014Butterfly,MaldacenaShenkerStanford2016ChaosBound}.
The spirit of this approach is reminiscent of, and inspired by, Zurek's  ``predictability sieve"  \cite{zurekPreferredStatesPredictability1993}.
Among a plethora of operationally available partitions into (generalized) subsystems, a partition is selected by the mereological criterion of minimal scrambling. This dynamically emergent partition is the one that is most resilient against information leakage and delocalization entailed by quantum dynamics  \cite{zanardiOperationalQuantumMereology2024,andreadakisLongtimeQuantumScrambling2024}.

The question that is addressed in this paper is: how sensitive is this particular mereological choice with respect  to a change of Hamiltonian parameters?

To answer this question we will pursue a strategy  analogous to the differential-geometric --so-called fidelity--  approach to  Quantum Phase Transitions (QPT) \cite{Gu2010FidelityReview,zanardiGroundStateOverlap2006,zanardiInformationTheoreticDifferentialGeometry2007}.
The generalized subsystems will be associated to operator subalgebras or, equivalently, to the corresponding projection CP-maps.
Over these projections, a natural distance function is defined and its infinitesimal form leads to a Riemannian metric. Scrambling minimization induces a map between the Hamiltonian parameter space 
and the manifold of operator algebras (projections)  which can be used to pull the metric back.  

We will show that the pulled-back metric can, in the limit of large system size,  exhibit singularities analog to those encountered in ordinary QPT's.
We dub this phenomenon a {\em{mereological phase transition}} (m-QPT). At these points, a small change of Hamiltonian leads to a large change of the  dynamically emergent, i.e., minimally scrambling, decomposition into subsystem.
 We will present analytical results with toy models as well numerical ones for many-body systems featuring chaos-integrability and localization transitions.

{{\prlsection{$g$-TPS and scrambling}}}
We begin with a family $\mathcal{A} \equiv \{ A_\theta \}_{{\theta}\in\Theta}$ of isomorphic $*$-subalgebras of $\mathcal{B}(\mathcal{H})$, where $\mathcal{H}$ is a finite $d-$dimensional Hilbert space. Here, $\Theta$ denotes a manifold
of parameters and the map $\theta\mapsto A_\theta$ is injective.  Intuitively, the $A_\theta$'s  correspond to different operational resources available to the observer, e.g. alternative setups of implementable quantum operations. 

Following the perspective of quantum mereology  explored in Refs. \cite{carrollQuantumMereologyFactorizing2021,zanardiOperationalQuantumMereology2024,andreadakisLongtimeQuantumScrambling2024}, each of the $A_\theta$'s defines a generalized  tensor product structure (g-TPS).  
Indeed, for fixed $\theta$, the Hilbert space breaks down into a direct sum of virtual bi-partitions \cite{zanardiVirtualQuantumSubsystems2001,zanardiQuantumTensorProduct2004}:
\begin{equation}\label{eq:H-decomp}
\mathcal{H}\cong_\theta \bigoplus_J\,  \mathbf{C}^{n_J} \otimes \mathbf{C}^{d_J}.
\end{equation}
In each of the $J$-blocks, the observer acts irreducibly on the second factor and trivially on the first one by implementing elements of $A_\theta.$
Quantum information can be encoded and manipulated in the $ \mathbf{C}^{d_J}$, while the  $\mathbf{C}^{n_J}$ describe an effective environment with inaccessible degrees of freedom.  If $A_\theta$ is replaced by its {\em{commutant}} $A_\theta^\prime:=\{ X\in \mathcal{B}(\mathcal{H})\colon [X, A_\theta]=0\}$ the role of the factors is inverted.    
The maximal projections of the center $Z_\theta:=A_\theta\cap A_\theta^\prime$ may be thought of as superselection charges and correspond to the $J$ labels. In the following, g-TPS refers interchangeably to the $A_\theta$'s or the associated decomposition (\ref{eq:H-decomp}).
The algebra $A_\theta$ is maximal abelian (``factor'') if $|\{J\}|=d$  ($|\{J\}|=1$). The first case corresponds to the selection of a complete set of commuting observables the second to a  bipartition.

Next, we consider a Hamiltonian $H(\lambda)\in \mathcal{B}(\mathcal{H})$, where $\lambda\in\mathcal{M}$ denotes physical parameters, e.g.  coupling constants or external fields.  These are the dynamical inputs in our approach and will be used  to select a distinguished g-TPS from the family $\mathcal{A}$ of potentially available ones. 
 The quantum evolution will generically disrupt decomposition (\ref{eq:H-decomp})  by allowing information leakage  between the different $J$-blocks as well as by generating quantum entanglement between the virtual subsystems.

In order to measure this disruption we will now introduce 
a third ingredient:  a real-valued scrambling functional
$\sigma(A_\theta, H(\lambda)).$
This is an entropy-like  measure of the scrambling of the g-TPS associated with $A_\theta$ induced by the unitary dynamics generated by $H(\lambda).$
Qualitatively, $\sigma$ assesses the stability of the g-TPS against the delocalization of  quantum information entailed by the dynamics.  The lower $\sigma$, the more stable the g-TPS.

We are finally in the position of bringing these different ingredients together and to outline a key step of our strategy. For {\em{fixed}} Hamiltonian parameters $\lambda$, we define
\begin{equation}\label{eq:theta_min}
\theta_{\min}(\lambda) = \arg \min_{{\theta}\in\Theta} \sigma(A_\theta, H(\lambda))\in\Theta.
\end{equation}
Assuming that this minimum is unique, this
yields a map ${\mathcal{M}}\rightarrow\Theta\rightarrow{\mathcal{A}}\colon\lambda \mapsto A_{\theta_{\min}}.$ This map selects the most  resilient g-TPS (as quantified by $\sigma$) in $\mathcal{A}.$
Following the approach to  quantum mereology advocated in Refs. \cite{zanardiOperationalQuantumMereology2024,andreadakisLongtimeQuantumScrambling2024}, we regard this minimally scrambling g-TPS as the naturally emergent one.

To   introduce our  concrete choices for the scrambling functional, we first define the so-called $A$-OTOC associated to an observable algebra $A$ and a unitary dynamics $U_t=e^{i t H}$ by \cite{andreadakisScramblingAlgebrasOpen2023}
\begin{equation}\label{eq:A-OTOC}
G_A(t)=\frac{1}{2d}\mathbb{E}_{X,Y}\left[\|[X, \mathrm{Ad} U_t (Y) ]\|_2^2\right],
\end{equation}
where the expectation is over Haar-distributed unitaries $X\in A$ and $Y\in A^\prime$ and $ \mathrm{Ad} U(Y):=U Y U^\dagger$ is the adjoint action of $U.$
The norm is the standard operator Hilbert-Schmidt one $\|X\|_2:=\sqrt{\mathrm{Tr} X^\dagger X}.$

The A-OTOC, \cref{eq:A-OTOC},  for different choices of $A,$ corresponds, e.g., to coherence-generating power (for a maximal abelian subalgebra) and operator entanglement (for a factor)  \cite{yanInformationScramblingLoschmidt2020,styliarisInformationScramblingBipartitions2021,zanardiQuantumScramblingObservable2022,andreadakisScramblingAlgebrasOpen2023}.
The A-OTOC would seem to provide a first natural choice for a scrambling functional. However its explicit dependence on finite-time evolution makes it very hard to handle in non-trivial situations.

Therefore, in this paper we will instead adopt as physically motivated proxies its short-time expansion  \cite{zanardiOperationalQuantumMereology2024}  and  long time average \cite{andreadakisLongtimeQuantumScrambling2024}.
In Ref. \cite{zanardiOperationalQuantumMereology2024}, it was shown that the short-time expansion of \cref{eq:A-OTOC} reads: $G_A(t)= \frac{2}{d} \,\sigma_s^2\,t^2 +O(t^3)$, where 
\begin{equation}\label{eq:sigma-short}
\sigma_s(A_\theta, H):=\|Q_\theta(H)\|_2
\end{equation}
and $Q_\theta:=1-P_{A_\theta + A^\prime_\theta}$ is the projection CP-map onto the orthogonal complement of the subspace
$A_\theta + A^\prime_\theta\subset \mathcal{B}(\mathcal{H}).$
\cref{eq:sigma-short} provides our first choice for the scrambling functional.  It  describes the (Gaussian) rate at which the A-OTOC, \cref{eq:A-OTOC}, grows at short times. The smaller the $\sigma_s$, the slower the degrees of freedom associated with $A$ will start scrambling with those associated to $A^\prime.$

A natural and complementary choice for $\sigma$ is given by the long time behavior of the A-OTOC:
\begin{equation}\label{eq:sigma-long}
\sigma_l(A,H)=\lim_{T\to\infty}\frac{1}{T}\int_0^T G_A(t) \,dt.
\end{equation}
%
Physically,  the smaller the $\sigma_l$, the lesser is  information scrambling between $A$ and $A^\prime$ observed over a sufficiently long time-scale.  This quantity,  when $A$ is a factor,
has been shown to be able to effectively tell apart quantum chaotic Hamiltonians from integrable ones and those giving rise to (many-body)localization \cite{styliarisInformationScramblingBipartitions2021,anandBROTOCsQuantumInformation2022}.

\begin{figure}[t]
\centering
\framebox{
\begin{minipage}{0.47\textwidth}
\small
\[
\begin{array}{|c|c|}
\hline
\textbf{Standard QPT} & \textbf{Mereological QPT} \\
\hline
H(\lambda) & H(\lambda) \\
\{ \ket{\psi_\theta} \} & \{ A_\theta \} \\
E(\theta) = \bra{\psi_\theta} H(\lambda) \ket{\psi_\theta} & \sigma(A_\theta, H(\lambda)) \\
\theta_{\min} = \arg \min_\theta E & \theta_{\min} = \arg \min_\theta \sigma \\
ds^2 = \braket{d\psi_\theta | d\psi_\theta} 
- |\braket{\psi_\theta | d\psi_\theta}|^2 & 
ds^2 = \| dP_{A_\theta} \|_{\mathrm{HS}}^2 \\
\hline
\end{array}
\]
\end{minipage}
}
\caption{ \label{fig:analogy} Conceptual analogy between standard and mereological quantum phase transitions (QPTs). The minimization over states is replaced by one over algebraic structures.}
\label{fig:qpt_vs_mqpt}
\end{figure}
\par \prlsection{Metric structure and $m$-QPTs}
The approach to m-QPTs explored in this paper parallels the differential-geometric, fidelity, approach to quantum phase transitions \cite{zanardiGroundStateOverlap2006,Gu2010FidelityReview,zanardiInformationTheoreticDifferentialGeometry2007}. There, the singularities 
of the pull-back of the Fubini-Study metric on Hamiltonian parameter spaces identify critical points. Specifically, the present analogy consists in  replacing states and energy minimization with algebras and scrambling minimization respectively (see \cref{fig:analogy}).

A metric structure over the g-TPS family  $\mathcal{A}$  is now required to make the analogy complete. To construct one, let us first observe that
 $\mathcal{A}$  is a subset of the Grassmannian manifold of subspaces of  $\mathcal{B}(\mathcal{H}). $ This shows that $\mathcal{A}$  is naturally  equipped with a distance function:
$D(A_{\theta'}, A_\theta) =d^{-1} \| P_{{\theta'}} - P_{\theta} \|_{\mathrm{HS}},$
where $P_\theta$ is the projection CP-map onto $A_\theta$, $\| X\|_{\mathrm{HS}}:=\sqrt{\langle X, X\rangle_{\mathrm{HS}}}$ denotes the Hilbert-Schmidt norm of maps 
\footnote{If $T$ and $F$  are CP-maps we define $\langle T, F\rangle_{\mathrm{HS}}:= \sum_{l,m}  \langle T( |l\rangle\langle m|),  F( |l \rangle\langle m|)\rangle,$ where the kets denote an orthonomal basis of $\cal H$ and the scalar product is standard Hilbert-Schmidt one for operators 
i.e., $\langle A, B\rangle=\mathrm{Tr}( A^\dagger B)$. }.
It is convenient to map the g-TPS's $A_\theta$ onto a manifold of quantum states. By introducing  the Choi representation  \cite{Choi1975}   $P_\theta\rightarrow\rho_\theta:=(P_\theta\otimes\mathbf{1})(|\Phi^+\rangle\langle\Phi^+|)$, $|\Phi^+\rangle=d^{-1/2} \sum_{i=1}^d |i \rangle^{\otimes 2}$, one finds
\begin{equation}\label{eq:dist-HS}
D(A_{\theta'}, A_\theta) =\|\rho_{{\theta'}} - \rho_{\theta} \|_2.
\end{equation}
We notice that,  $D(A_{\theta'}, A_\theta) \le \|\rho_{{\theta'}} - \rho_{\theta} \|_1\le 2 D_B(\rho_\theta,\rho_{\theta'}),$ where $D_B$ is the Bures distance \cite{Bures1969}. 
Also, $D(A_{\theta'}, A_\theta) \le \DiamondNorm{P_\theta-P_{\theta'}}$ where $\DiamondNorm{\cdot}$ is the diamond norm \cite {Watrous2009CBnormSDP}.
These inequalities  show that  the  distance (\ref{eq:dist-HS}) bears  non-trivial  information-theoretic content.  The larger the distance, the  higher is the degree of statistical distinguishability between the states (CP-maps)
$\rho_\theta$ and $\rho_{\theta'}$ ($P_\theta $ and $P_{\theta'}$). 

By considering infinitesimally close g-TPSs, and thereby $A_\theta$'s,  \cref{eq:dist-HS} can be used to define  a Riemannian metric over $\Theta$:
$ds_\Theta^2 =D(A_{\theta'},  A_{\theta+d\theta})^2= \sum_{\mu,\nu=1}^{\mathrm{dim}{\Theta} }  g_{\mu\nu}^\Theta d\theta^\mu d\theta^\nu.$
Finally,  by using \cref{eq:theta_min} this metric  can be  pulled back to  $g^{\cal{M}}$ on the parameter space $\mathcal{M}.$ 
The tensors $g^\Theta$ and $g^{\cal{M}}$ will be sometimes in the following --in analogy to the ordinary QPT case-- referred  to as {\em{algebra susceptibilities}}.
Note also that the following bound holds  $ds_\Theta^2 \le 4 ds_B^2$, where $ds_B^2$ is the Bures metric element \cite{Bures1969, Uhlmann1976}, which is proportional to the quantum Fisher information metric on the manifold  $\{\rho_\theta\}_{\theta\in\Theta}$ \cite{BraunsteinCaves1994}.


Having  established a full parallelism with the ingredients of the metric approach to QPTs, we can identify
\textit{mereological quantum phase transitions} (m-QPTs) by the singularities of $g^{\mathcal{M}}$ in the thermodynamic limit.
The intuition behind this definition is that at an m-QPT, a small change $\lambda\rightarrow \lambda+d\lambda$ in the  parameters of the Hamiltonian gives rise  to a large change in the associated  g-TPS's  $A_{\theta_{\min}(\lambda)}$ and $A_{\theta_{\min}(\lambda+\delta\lambda)}.$ 

In the rest of this  paper, we will focus on the case where each subalgebra is unitarily  and uniquely related to a fixed reference: 
$A_\theta = \mathrm{Ad}\, U_\theta (A_0),\,(\theta\in\Theta), $
where the family of unitaries $U_\theta$ is closed under hermitian conjugation.
The scrambling functionals based on the A-OTOC (\ref{eq:A-OTOC}) have the covariance property
\begin{equation}\label{eq:covariance}
\sigma(  A_\theta, H)= \sigma(A_0, \mathrm{Ad}\, U^\dagger_\theta (H))\quad(\forall\theta\in\Theta).
\end{equation}
From this it follows that the minimization (\ref{eq:theta_min}) can be equivalently thought of as being performed over a family of rotated Hamiltonians
 with respect to the  {\em{fixed}} g-TPS associated with $A_0.$  

In order to illustrate our  method, we will also  assume that the  algebra $A_0$ is one of the two following two types:
{\bf{1)}} The algebra of operators diagonal  in a given orthonormal basis of $\mathcal{H}$ or {\bf{2)}} the factor algebra  $\mathcal{B}(\mathcal{H}_A)\otimes \openone_B$ on a Hilbert space $\mathcal{H}=\mathcal{H}_A\otimes\mathcal{H}_B.$

In these cases the g-TPS distance (\ref{eq:dist-HS}) has a remarkably direct physical and operational meaning for unitary families. Namely, if $W:= :U_{\theta'}^{-1}U_\theta,$  then  Eq.~(\ref{eq:dist-HS}) is proportional to  the coherence-generating power \cite{ZanardiCoherenceGenerating2017,zanardiQuantumCoherenceGenerating2018} (case {\bf{1)}}) or to  the operator entanglement  \cite{zanardiEntanglementQuantumEvolutions2001,ProsenOperatorSpaceEntanglement2007} (case  {\bf{2)}}) of $W.$ 
Furthermore, it  can be shown  [See \cref{app:proof}] that the metric element  $ds^2_\Theta$  can be written as,  
\begin{equation}\label{eq:metric-Theta}
ds^2_\Theta= \kappa^2 \,\| Q_\theta(dK)\|_2^2,
\end{equation}
where $\kappa=2\,(\mathrm{dim} A_0')^{-1},$ and $dK= -i dU_\theta U^\dagger$ \footnote{
Since $Q_\theta$ is a projector one has that  $ds^2_\Theta\le \kappa ^2\| dK\|_2^2= \kappa^2 \|dU\|_2^2,$ and equality holds when $Q_\theta(dK)=dK.$}.

We will now construct a couple of toy models involving the cases 1) and 2) above which will allow us to show analytically how m-QPT's may arise.

\par \prlsection{Maximal abelian}
Consider case 1), the maximal abelian algebras where $A=A^\prime.$  Here, the g-TPS \cref{eq:H-decomp} consists of a direct sum of one-dimensional blocks and the minimally $\sigma_s$-scrambling g-TPS is obtained simply by making the (rotated) Hamiltonian diagonal.

Let $\mathcal{H} = (\mathbb{C}^2)^{\otimes N}$ be the space of $N$ qubits and the reference algebra $A_0 = \mathrm{span}\{\openone,  \sigma^z\}^{\otimes N}$. Set
$U_\theta = \bigotimes_{i=1}^N e^{i \theta_i \sigma^y / 2}$
and consider the Hamiltonian
\begin{equation}\label{eq:ham-ab}
H = \sum_{i=1}^N\left( \varepsilon_i \sigma_i^z + J_i \sigma_i^x\right).
\end{equation}
Here $\mathcal{M}$ is the manifold of the $\lambda:=(\{\varepsilon_i\}, \{J_i  \}).$
The projector CP-map $Q_0=1-P_{A_0}$ now projects to the operators which have just off-diagonal elements in the computational basis i.e., the joint eigenvectors of the $\sigma_i^z$'s.
In view of \cref{eq:covariance}, the scrambling rate (\ref{eq:sigma-short}) involves  just the $\sigma^x$-component of the rotated Hamiltonian
$H_\theta=\mathrm{Ad} U^\dagger_\theta(H). $  The minimization of $\sigma_s,$  which defines the function (\ref{eq:theta_min}), is then achieved when the  $\sigma^x$-component vanishes i.e., 
$\theta_i = \tan^{-1} \left(\frac{J_i}{\varepsilon_i}\right),\, (i=1,\ldots,N).$
 Since,  $Q_\theta(dK)=dK=\frac{1}{2}\sum_{i=1}^N d\theta_i \sigma_i^y$ one finds the induced metric (\ref{eq:metric-Theta})
\begin{equation}\label{eq:susceptibility}
ds_\mathcal{M}^2 \propto \sum_{i=1}^m d\theta_i^2 =\sum_{i=1}^N  \left(\frac{\varepsilon_i dJ_i-J_id\varepsilon_i}{\varepsilon_i^2 + J_i^2}\right)^2.
\end{equation}
The crucial observation is this expression is {\em{formally}} identical to the one for  fidelity susceptibility of the XY model \cite{zanardiInformationTheoreticDifferentialGeometry2007}. It then follows
that for some choices of the parameters $\lambda,$  corresponding to the ordinary QPT in that model,  one can observe a {\em{super-extensive}}  behavior of  $ds_\mathcal{M}^2$
for large $N$.  For example,  if $dJ_i=0\,(\forall i)$ and $J_i=O(1/N)$ and $\varepsilon_i=O(1/N^2)$ one sees that for $ N\to\infty$ the $i$-th term in the sum (\ref{eq:susceptibility})
is $O(1/J_i^{2})=O(N^2)$ which leads to a singular behavior. 
 \par \prlsection{Factor}
 We now move to a non-abelian (factor) algebra case. Here, minimizing $\sigma_s$ amounts to minimizing the interaction part of the (rotated) Hamiltonian witch generates entanglement between the subsystems.
Consider the bipartite space $\mathcal{H}=\mathcal{H}_L\otimes  \mathcal{H}_R$, where $\mathcal{H}_{L,R}=(\mathbf{C}^2)^{\otimes N}$ are $N$-qubits subsystems each.
The reference algebra is given by $A_0= \mathcal{B}(\mathcal{H}_L)\otimes\openone_R$ and $U_\theta=\otimes_{i=1}^N e^{i\frac{\theta_i}{2}  K_i^y},$ where $ K_i^y:= -i (
 \sigma_{iL}^+\sigma_{iR}^-- \sigma_{iR}^+\sigma_{iL}^- ).$
 The projection CP-map $Q_0=1-P_{A_0+A_0^\prime}$ now singles out the non-trivial couplings between  $\mathcal{H}_L$ and $ \mathcal{H}_R.$
 The Hamiltonian is given by $H=\sum_{i=1}^N H_i$ where
 \begin{equation}
 H_i= \varepsilon_{iL}\sigma_{iL}^z +      \varepsilon_{iR} \sigma_{i R}^z +J_i(\sigma_{iL}^+\sigma_{iR}^-+   \sigma_{iR}^+\sigma_{iL}^-  ).
 \end{equation}
 We choose for simplicity $\varepsilon_{iL}= - \varepsilon_{iR}=\varepsilon_i.$ Now notice that the Hamiltonians $H_i$ are trivial on $\mathrm{span}\{ |00\rangle,\,|11\rangle\}\subset \mathbf{C}^2_{iL}\otimes \mathbf{C}^2_{iR}$ and identical to (\ref{eq:ham-ab}) on $\mathrm{span}\{ |01\rangle,\,|10\rangle\}\subset \mathbf{C}^2_{iL}\otimes \mathbf{C}^2_{iR}.$ 
 Since the same is true for the generator $dK=Q_0(dK)=\sum_{i=1}^N d\theta_i K^y_i$ one reaches the conclusion that the metric structure of this problem is isomorphic to the one of the former maximal abelian one.  This shows that this case features a m-QPT as well.
\vskip 0.2truecm
A a couple of remarks, valid for both  examples above,  are now in order.  
a) The scrambling function $\sigma_s$ itself remains {\em{regular}} at the m-QPT ($\sigma_s( A_{\theta_{\min}(\lambda)}, H(\lambda))=0$ as a function of $\lambda$). 
This is in contrast with what happens in ordinary QPTs, where the energy function has singularities at the critical points.
b) Since the metric $g^\Theta$ is flat i.e.,  $\theta$-independent, the behavior -- regular or otherwise -- of  $g^{\cal M}$ is controlled entirely by the function (\ref{eq:theta_min}). This is not true in the general case.

 \par \prlsection{m-QPTs in  Many-body Systems }
 To further establish the relevance of m-QPTs in quantum many-body transitions, we numerically investigate quantum many-body Hamiltonians across regimes with distinct quantum many-body properties i.e., chaotic and integrable. This change in integrability properties is associated with significant changes in the eigenstate structure \cite{albaEigenstateThermalizationHypothesis2015,khassehIdentifyingQuantumManyBody2023}. 
 
 Motivated by this, we analyze the behavior of the  long-time  scrambling functional $\sigma_l$, \cref{eq:sigma-long}, across an integrability transition for the case where ${A}_0$ is the factor subalgebra for a bipartition of a quantum spin chain and the family of unitaries is the whole unitary group. 
 
 In Ref. \cite{andreadakisLongtimeQuantumScrambling2024},  it was shown that the minimization of this quantity is closely related to the eigenstate entanglement properties of the Hamiltonian:
{the smallest $\sigma_l$ is achieved when the amount of entanglement in the eigenstates of the Hamiltonian (with respect to the given partition) is the smallest.} Therefore, from the dual point of view (\ref{eq:covariance}), our optimization problem can be intuitively (and roughly) cast as: given a fixed bipartition and a (unitary) family of Hamiltonians find the one whose eigenstates have the minimum amount of entanglement.
 
\begin{figure}
\centering
\includegraphics[width=\columnwidth]{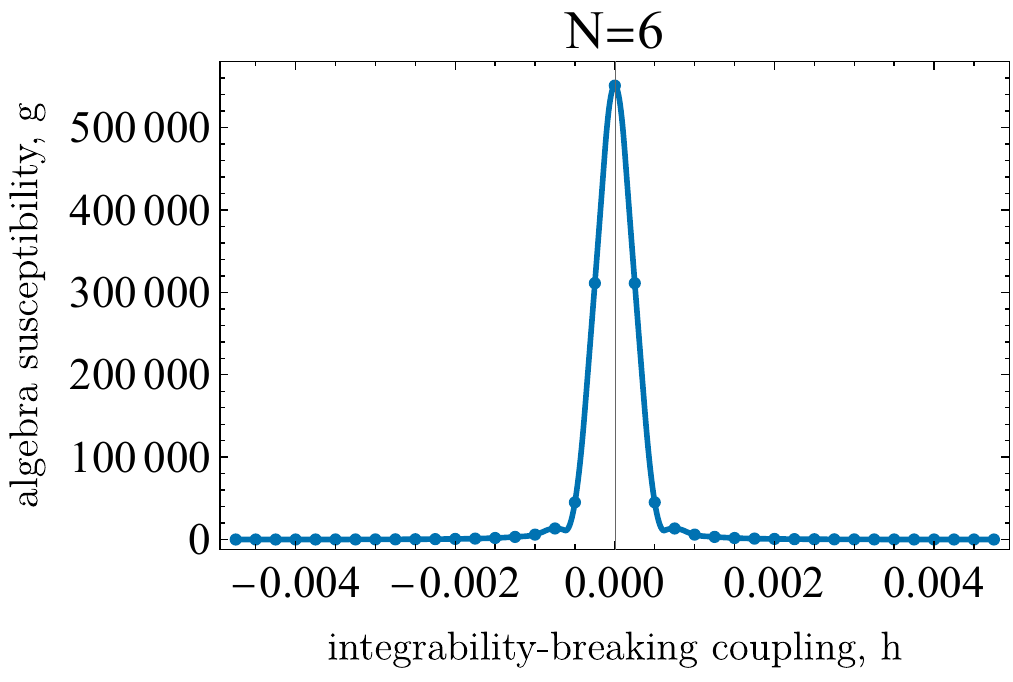}
\caption{The algebra susceptibility $g$ for the minimally scrambling symmetric bipartitions of the transverse field Ising model for various points in the Hamiltonian parameter space $h$. For $h=0$, the model can be mapped to free fermions and is nonintegrable otherwise. We observe that at the point of the integrability transition, the algebra susceptibility becomes sharply peaked, indicating the onset of an m-QPT.}
\label{fig:int}
\end{figure}

More specifically, we consider as a prototypical toy-model the well-studied transverse field Ising model with on-site magnetization in one dimension:
 \begin{equation} \label{tfim}
H(h,g)=-\sum_{i=1}^{N-1} \sigma_i^z \sigma_{i+1}^z - \sum_{i=1}^N (h \sigma_i^z + J \sigma_i^x).
 \end{equation}
For $J \neq 0$, this model can be mapped to free fermions via a Jordan-Wigner transformation \cite{liebTwoSolubleModels1961,pfeutyOnedimensionalIsingModel1970} if $h=0$, while it is nonintegrable otherwise \cite{kimBallisticSpreadingEntanglement2013}. 
Fixing $J=1.05$, we use a gradient descent algorithm, see \cref{app:gradient}, to find a unitary rotation $U_{\text{min}}(h) = \arg \min_{U} \sigma_l(\mathrm{Ad} U ({A}), H(h,J=1.05))$ for small variations of the coupling strength $h$ around $h=0$.  This identifies, for every $h$, an isomorphic bipartite algebra ${A}_\alpha \coloneqq \mathrm{Ad} U_{\text{min}}(h_\alpha) ({A})$ for which the long time average of the ${A}$-OTOC is minimized.
Then, we numerically compute the algebra susceptibility \cref{eq:metric-Theta} for each point $h_\alpha$ in the Hamiltonian parameter space, which we plot in \cref{fig:int}. Further details about the numerical methods can be found in \cref{app:int}. The algebra susceptibility shows a sharp peak around $h=0$, which provides evidence for a m-QPT of the minimally scrambling bipartition, which is associated to the integrability transition.
\par In \cref{app:loc}, we discuss some further numerical observations for the case that the transverse field coupling is disordered. While some of the details of the method used there are different, it shows that the algebra susceptibility can, in a similar way, be sensitive to the ``emergent'' integrability associated with the transition to a many-body localized phase \cite{nandkishoreManyBodyLocalizationThermalization2015,imbrieLocalIntegralsMotion2017}.
\par \prlsection{Conclusion}
We have considered a  family of generalized tensor product structures (g-TPS) over the Hilbert space of a quantum system. 
Each of these g-TPS is defined by a hermitian closed subalgebra of operators whch induces a decomposition of the Hilbert space with an additive and a nested tensor structure.
Given a Hamiltonian and a quantitative measure of quantum scrambling one can identify the dynamically emergent g-TPS in the family  as the one which  is minimally scrambling.
In this way, the subsystem ``identity'' is maximally robust against information leakage.

The family of g-TPS is mappable to a manifold of quantum states in a doubled Hilbert space. This manifold is in turn endowed with a natural distance and a Riemannian metric. 
We have shown that for some critical values of the Hamiltonian parameters this metric develops singularities in the large system size limit. 

We have provided analytical examples of m-QPTs for toy models and presented numerical evidence that the m-QPT concept is relevant to the chaos-integrability and localization transition in physical many-body problems.

The ideas presented in this paper are speculative in nature and lie at the intersection of quantum foundations, quantum information theory and many body physics. 
Their full scope and their ultimate physical relevance are matter for future critical investigations.


\par \prlsection{Acknowledgements}
PZ acknowledges partial support from the NSF award
PHY2310227. FA acknowledges financial support from a University of
Southern California “Philomela Myronis” scholarship.
\bibliographystyle{apsrev4-2}
\bibliography{paper.bib}
\onecolumngrid
\newpage
\appendix
\setcounter{secnumdepth}{2}
\crefalias{section}{appendix}  
\section{Details for the integrability transition algorithm}\label{app:int}
Our strategy to obtain \cref{fig:int} is as follows. For a fixed system dimension $N=6$ and coupling $J=1.05$, we take $N_{\text{steps}}+1$ equidistant values $h_\alpha$ from the interval $[0,0.005]$, starting from $h=0$. As disucssed in the main text, for each of these values, we use the gradient descent algorithm described in \cref{app:gradient} to find a unitary rotation $U_{\text{min}}(h_\alpha) = \arg \min_{U} \sigma_l(\mathrm{Ad} U ({A}), H(h,J=1.05))$. Here, ${A}$ is chosen to be the subsystem algebra for the subsystem composed of the first $N/2$ qubits in the spin-chain. For every step after $h=0$, we use the previously identified unitary to initialize the search for $U_{\text{min}}$. This identifies, for every $h_\alpha$, an isomorphic bipartite algebra ${A}_\alpha \coloneqq \mathrm{Ad} U_{\text{min}}(h_\alpha) ({A})$ for which the long time average of the ${A}$-OTOC is minimized. For convenience, we repeat this process for the interval $[-0.005,0]$. Then, we numerically compute the algebra susceptibility, \cref{eq:metric-Theta}, as
\begin{equation} \label{eq:num_suscept}
g(h_\alpha)=\frac{D({A}_\alpha,{A}_{\alpha-1})^2+D({A}_\alpha,{A}_{\alpha+1})^2}{2\,\delta h^2}.
\end{equation}
Here $D({A}_\alpha,{A}_{\alpha+1})^2$ denotes the square of the algebra distance (\ref{eq:dist-HS}) and $\delta h = 0.005/N_{\text{steps}}$ is the numerical step of the sampling. It is worth noting that in this bipartite case, this squared distance is proportional to the operator entanglement of the unitary $U= U_\alpha \, U_{\alpha+1}^\dagger$ \cite{styliarisInformationScramblingBipartitions2021}.
\section{Gradient descent algorithm} \label{app:gradient}
In Ref. \cite{zanardiQuantumScramblingObservable2022},  under the condition of no resonances in the spectral gaps (NRC) it was shown that 
\begin{equation} \label{NRC}
        \overline{G_{{A}}(t)}^{NRC} = 1-\frac{1}{d}\left(\sum_{X=\{{A},{A}^\prime\}}\Tr( R^{(0),{X}}R^{(1),{X}^\prime} ) -\frac{1}{2}\Tr( R_D^{(0),{X}}R_D^{(1),{X}^\prime} )\right).
        \end{equation}
Here, for algebra ${X}$, $R_{lk}^{(0), {X}} \coloneqq \left\lVert \mathbb{P}_{{X}}\left(\ket{\phi_k}\bra{\phi_l} \right) \right\rVert_2^2$ and $R_{kl}^{(1),{X}} \coloneqq  \left\langle\mathbb{P}_{{X}}\left(\Pi_k\right),  \mathbb{P}_{{X}}\left(\Pi_l\right)\right\rangle$, where $\ket{\phi_k}$ denote the Hamiltonian eigenstates and $\Pi_k \coloneqq \ketbra{\phi_k}{\phi_k}$. In addition, for matrix $M$, $M_D \coloneqq diag(M)$. 
For the purposes of this section, ${A}$ is the subalgebra $\mathcal{B}(\mathcal{H}_A) \otimes \openone_B$ corresponding to a system bipartition $\mathcal{H} = \mathcal{H}_A \otimes \mathcal{H}_B$, in which case $\sqrt{\dim({X})} R_{lk}^{(0),{X}^\prime}= \sqrt{\dim({X}^\prime)} R_{kl}^{(1),{X}}=\langle \rho_k^X, \rho_l^X \rangle$ are the Gram matrices of the reduced Hamiltonian eigenstates $\rho_k^X$ on the X subsystem \cite{styliarisInformationScramblingBipartitions2021}. Hereafter, we consider the case that $N$ is even and fix $\mathcal{H}_A \cong \mathcal{H}_B \cong \mathbb{C}^{N/2}$.

The gradient descent algorithm is due to Abrudan \textit{et al.} \cite{abrudanSteepestDescentAlgorithms2008}. A quick exposition to the relevant details can be found in Appendix B of Ref. \cite{andreadakisOperatorSpaceEntangling2024}. For our purposes, given the bipartition $\mathcal{H} \cong \mathcal{H}_A \otimes \mathcal{H}_B \cong \mathbb{C}^{N/2} \otimes \mathbb{C}^{N/2}$ and a Hamiltonian $H$, we search for a unitary $U_{\text{min}}$, such that the unitarily transformed bipartition minimizes the quantity in \cref{NRC}. Equivalently, due to the duality (\ref{eq:covariance}) between unitary rotations of the algebra and the Hamiltonian, we can instead search for the corresponding unitary rotation $V_\text{min} \equiv U_{\text{min}}^\dagger$ of the Hamiltonian, where we have introduced additional notation to avoid confusion.
\par The function to be minimized, \cref{NRC}, in the bipartite case takes the form \cite{styliarisInformationScramblingBipartitions2021}
\begin{equation}
    f \coloneqq \overline{G_{\mathrm{Ad}V^\dagger ({A})}(t)}^{NRC} = 1-\frac{1}{d^2} \sum_{k,l=1}^d \sum_{X=\{A,B\}} \left(1- \frac{\delta_{kl}}{2}\right) \left( \langle V \rho_k^X V^\dagger, V \rho_l^X V^\dagger \rangle^2 + \langle V \rho_k^{\bar{X}} V^\dagger, V \rho_l^{\bar{X}} V^\dagger \rangle^2 \right).
\end{equation}
Starting from an initializing unitary $V^0$, the direction of steepest ascent is given by $G_V \coloneqq \Gamma_V \, V^\dagger - V \Gamma_V^\dagger$, where $\delta f = 2 \mathscr{R} \Tr(\Gamma_V^\dagger \delta V)$ for a variation $\delta V$ \cite{abrudanSteepestDescentAlgorithms2008,andreadakisOperatorSpaceEntangling2024}. Performing the variation, and after some algebra, we get
\begin{equation}
\Gamma_V= -\frac{4}{d^2} \sum_{k,l=1}^d \sum_{X=\{A,B\}} \left( 1-\frac{\delta_{kl}}{2}\right) \left(\Tr_{X^\prime , \bar{X}^\prime}(S_{XX^\prime} V\Pi_k \otimes V\Pi_l V^\dagger ) + \Tr_{X^\prime , \bar{X}^\prime}(S_{\bar{X} \bar{X}^\prime} V\Pi_k \otimes V\Pi_l V^\dagger) \right), 
\end{equation}
where $S_{XX^\prime}$ is the swap operator between the subsystem $X$ and its copy $X^\prime$ in the doubled Hilbert space $\mathcal{H}^{\otimes 2}$. Then, we iteratively update the unitary $V^{k+1}=\exp(-\mu_k G_k) V^k$, where $\mu_k$ is a dynamically adjusted step size to increase efficiency \cite{abrudanSteepestDescentAlgorithms2008}. We stop the search when the convergence condition $\lvert f(V^{k+1}) - f(V^k)\rvert \leq \epsilon \coloneqq 10^{-10}$ is met.
\par For the purposes of \cref{fig:int}, we initialize the unitary search with $V^0=\openone$ for the first point in parameter space ($h=0$), while we use the result of the last search for the other points in parameter space.
\section{Algebra susceptibility and the localization transition} \label{app:loc}
\begin{figure}
\begin{subfigure}{0.55\textwidth}
\centering
\includegraphics[width=1\linewidth]{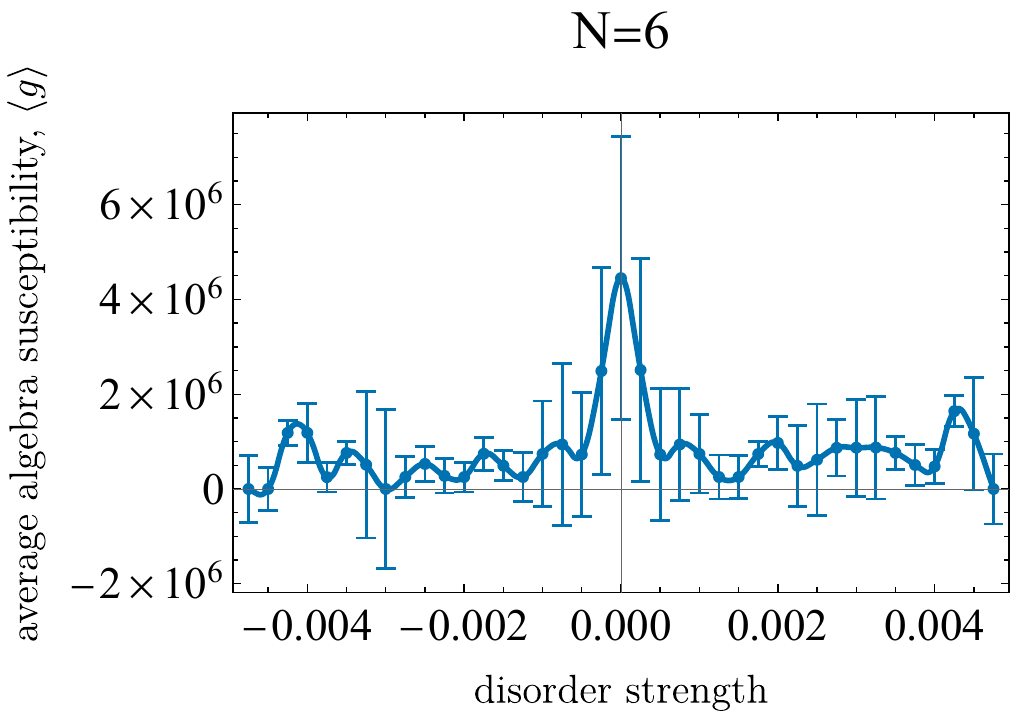}
\caption{}
\label{fig:loc}
\end{subfigure}%
\begin{subfigure}{0.45\textwidth}
\centering
\includegraphics[width=1\linewidth]{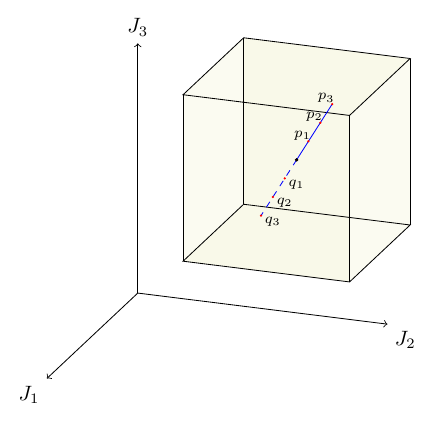}
\caption{}
\label{fig:graphic}
\end{subfigure}%
\caption{(a) The algebra susceptibility for the minimally scrambling symmetric bipartitions for the disordered transverse field Ising model using the method discussed in \cref{app:loc}. Zero disorder corresponds to the nonintegrable model, while otherwise the model is in a localized phase. We observe that the algebra susceptibility grows by an order of magnitude at the point of this localization transition, which can be seen as a response to the ``emergent'' integrability of the localized phase. (b) Visual representation of the parameter points sampling for the disordered model for $N=3$ and $N_\text{steps}=3$. The first point, $p_3$ is drawn randomly within the shaded cube. Then, the points $p_2$ and $p_1$ are chosen equidistantly on the line that connects $p_3$ and the center of the cube that represents the zero disorder point. The points $q_1$, $q_2$ and $q_3$ are then the symmetrically opposite points of increasing disorder strength. }
\end{figure}
Assume that the transverse field couplings in \cref{tfim} are now site-dependent
 \begin{equation} \label{tfim_dis}
H(h,g)=-\sum_{i=1}^{N-1} \sigma_i^z \sigma_{i+1}^z - \sum_{i=1}^N (h \sigma_i^z + J_i \sigma_i^x).
 \end{equation}
For a fixed $h \neq 0$, any amount of disorder in the transverse field coupling $J_i$ gives rise to a localized phase known as many-body localization \cite{nandkishoreManyBodyLocalizationThermalization2015}, which can be understood as an ``emergent'' integrability arising from an extensive number of local integrals of motion \cite{imbrieLocalIntegralsMotion2017}. Here, we report some numerical results that indicate that the algebra susceptibility (\ref{eq:metric-Theta}) is also sensitive to this transition.
\par The method we use is similar to the one used in the main text for the integrability transition. Here, we fix $h=0.5$ and draw each $J_i$ uniformly from $[1.05-\delta,1.05+\delta]$, where $\delta=0.005$ is a chosen maximum disorder strength, which is proportional to the standard deviation of this probability distribution. Then, we generate $N_{\text{steps}}+1$ sets of $J_i$ by sampling equidistantly the line that connects the initially drawn set and the nonintegrable point $J_i=J=1.05$ in the parameter space $\mathbb{R}^N$, see \cref{fig:graphic}. Notice that, effectively, this gives a single-shot process of decreasing disorder strength $\Delta_j = j/N_{\text{steps}} \, \delta, \; j=0,1,\dots,N_{\text{steps}}$, since the standard deviation of the random couplings is proportionally contracted. For every point in the parameter space, we use the gradient descent algorithm, \cref{app:gradient}, for $N=6$, to associate each disorder strength with a unitary $U_{\text{min}}(\{J_i\})$ and hence a bipartition ${A}(\{J_i\})$. For convenience, we extend this process for the symmetrically opposite points of increasing disorder strength (\cref{fig:graphic}). The difference here, compared to the case of the integrability transition discussed in the main text, is that at each point we initialize the search at $U^0=\openone$. Then, we use \cref{eq:num_suscept}, where $\delta h$ is replaced by the distance between the equidistant points in parameter space, to numerically determine the algebra susceptibility for each disorder strength and perform the disorder average over $n=5$ repetitions. The results are plotted in \cref{fig:loc}, where we used a negative disorder strength to simply denote the symmetrically opposite points in the parameter space. Once again, the algebra susceptibility grows significantly around the point $J_i=J=1.05$, which is where the transition between the localized and nonintegrable phase occurs.
\section{Derivation of Eq.~(\ref{eq:metric-Theta})}\label{app:proof}

Before proceeding to the proof  we must recall several  facts from Refs. \cite{zanardiQuantumScramblingObservable2022,andreadakisScramblingAlgebrasOpen2023}
\[
A=\operatorname{span}\{e_\alpha\}_{\alpha=1}^{d_0}
   =\operatorname{span}\{\tilde e_\alpha\}_{\alpha=1}^{d_0}
   \cong \bigoplus_{J}\, \mathbf 1_{n_J}\otimes M_{d_J}(\mathbb C);
\] where, from Eq.~(\ref{eq:H-decomp}),
\[
e_\alpha=\frac{1}{\sqrt{d_J}}\;\mathbf 1_{n_J}\otimes \bigl|\ell\bigr\rangle\!\bigl\langle m\bigr|, 
\qquad 
\tilde e_\alpha=\sqrt{\frac{d_J}{n_J}}\;e_\alpha;
\qquad 
\alpha=(J,\ell,m),\; J=1,\ldots,d_Z,\; \ell,m=1,\ldots,d_J.
\] and 
\[
A'=\operatorname{span}\{f_\beta\}_{\beta=1}^{d_0'}
   =\operatorname{span}\{\tilde f_\beta\}_{\beta=1}^{d_0'}
   \cong \bigoplus_{J}\, M_{n_J}(\mathbb C)\otimes \mathbf 1_{d_J};
\]
\[
f_\beta=\frac{1}{\sqrt{n_J}}\; \bigl|\ell\bigr\rangle\!\bigl\langle m\bigr| \otimes \mathbf 1_{d_J}, 
\qquad 
\tilde f_\beta=\sqrt{\frac{n_J}{d_J}}\;f_\beta;
\qquad 
\beta=(J,\ell,m),\; J=1,\ldots,d_Z,\; \ell,m=1,\ldots,n_J.
\]
These algebra bases
$
\{e_\alpha\}_\alpha, \{f_\beta\}_\beta \ \text{are orthogonal, and}  
\{\tilde e_\alpha\}_\alpha, \{\tilde f_\beta\}_\beta \ \text{are orthonormal.}
$
All are Hermitian closed. 

Collinear algebras are defined by the condition that $n_J/d_J$ is $J$-independent. In this case, this ratio equals $\sqrt{d_0'/d_0},$ where $d_0'=\mathrm{dim}\,A_0',$ and, $\mathrm{dim}\, {\cal B}({\cal H})=d^2=d_0\,d_0'.$ Moreover,
\[
\tilde e_\alpha=\sqrt{\frac{d_0}{d_0'}}\,e_\alpha, 
\qquad 
\tilde f_\beta=\sqrt{\frac{d_0'}{d_0}}\,f_\beta \qquad \mathrm{(collinear)}.
\]
In terms of these bases, we can express the algebra CP-map projections as follows:
\begin{equation}\label{eq:properties}
P_A(\,\cdot\,)=\sum_{\beta=1}^{d_0'} f_\beta (\,\cdot\,) f_\beta^\dagger,
\qquad
P_{A'}(\,\cdot\,)=\sum_{\alpha=1}^{d_0} e_\alpha (\,\cdot\,) e_\alpha^\dagger,
\qquad
\sum_{\alpha=1}^{d_0} e_\alpha^\dagger e_\alpha
=
\sum_{\beta=1}^{d_0'} f_\beta^\dagger f_\beta
= \mathbf 1 .
\end{equation}
It is also useful to introduce the following operators:
\[
\Omega_A := \sum_{\alpha=1}^{d_0} e_\alpha\otimes e_\alpha^\dagger,
\quad
\Omega_{A'} := \sum_{\beta=1}^{d_0'}  f_\beta \otimes  f_\beta^\dagger;
\qquad
\tilde\Omega_A = \sum_{\alpha=1}^{d_0} \tilde e_\alpha \otimes \tilde e_\alpha^\dagger,
\quad
\tilde\Omega_{A'} = \sum_{\beta=1}^{d_0'} \tilde f_\beta \otimes \tilde f_\beta^\dagger.
\]
One can move from $A$ to $A'$ by using a swap $S,$
\[
S\,\tilde\Omega_A=\Omega_{A'}, \qquad
S\,\tilde\Omega_{A'}=\Omega_A. 
\]
For the collinear case only,
\[
\tilde\Omega_A=\frac{d_0}{d_0'}\,\Omega_A,
\qquad
\tilde\Omega_{A'}=\frac{d_0'}{d_0}\,\Omega_{A'} .
\]
\subsection{}
If $A_\theta= \mathcal{U}(A_0)$ , where  $\mathcal{U}_\theta=\mathrm{Ad}\,{U}_\theta,$ and  $ P_0$ is the CP-projection onto  $A_0,$
then the CP-projection onto $A_\theta$ is given by:
$P_\theta \;=\; \mathcal{U}_\theta P_0 \mathcal{U}_\theta^\dagger,$ therefore

\[
\mathrm{d}P_\theta=\mathrm{d}\mathcal{U}_\theta\,P_0\,\mathcal{U}_\theta^\dagger+\mathcal{U}_\theta P_0\,\mathrm{d}\mathcal{U}_\theta^\dagger
=(\mathrm{d}\mathcal{U}_\theta \mathcal{U}_\theta^\dagger)P_\theta+P_\theta(\mathrm{d}\mathcal{U}_\theta \mathcal{U}_\theta^\dagger).
\]
Here we have used \(\mathrm{d}(\mathcal{U}_\theta \mathcal{U}_\theta^\dagger)=\mathrm{d}\mathcal{U}_\theta\,\mathcal{U}_\theta^\dagger+\mathcal{U}_\theta\,\mathrm{d}\mathcal{U}_\theta^\dagger
=i\,[\mathcal{K},P_\theta],\) where \(\mathcal{K}:=-i\,\mathrm{d}\mathcal{U}_\theta \mathcal{U}_\theta^\dagger=:\mathrm{ad}\, K,$ in which
$K:= -idU_\theta U^\dagger$ is an hermitian operator [{\em{Nota bene}}: in the main text it denoted by $dK.$] 

 \[
\|\,\mathrm{d}P_\theta\,\|_{\mathrm{HS}}^2
=\langle [\mathcal{K},P_\theta],[\mathcal{K},P_\theta]\rangle_{\mathrm{HS}}
=-\Tr_{\mathrm{HS}}\!\left\{\mathcal{K} P_\theta \mathcal{K} P_\theta + P_\theta \mathcal{K} P_\theta \mathcal{K} - \mathcal{K} P_\theta \mathcal{K} - P_\theta \mathcal{K}^2\right\}
\]
\[
=2\,\Tr_{\mathrm{HS}}\!\left(-\mathcal{K} P_\theta \mathcal{K} P_\theta + P_\theta \mathcal{K}^2\right)=2\,\left\|(\mathbf 1-P_\theta)\mathcal{K} P_\theta\right\|_{\mathrm{HS}}^{2}.
\]
Now let \(P_\theta=\sum_{\alpha=1}^{d_0}\ket{\alpha}\!\bra{\alpha}\), where \(\{\ket{\alpha}\}\) is an orthonormal basis of \(A_\theta\) and \(\dim(A_0)=d_0\). We have
\begin{equation}
\label{eq:I-II}
\frac{1}{2}\,\|\,\mathrm{d}P_\theta\,\|_{\mathrm{HS}}^2
=\sum_{\alpha=1}^{d_0}\left\|(\mathbf 1-P_\theta)\mathcal{K}\ket{\alpha}\right\|_2^{2}=\sum_{\alpha=1}^{d_0}\Bigl(\,\|\mathcal{K}\ket{\alpha}\|_2^{2}-
\sum_{\beta=1}^{d_0}|\bra{\beta}\mathcal{K}\ket{\alpha}|_2^{2}\Bigr)\equiv\mathrm{ {\bf{I-II}}}.
\end{equation} 
Now we evaluate these two terms separately.
\subsection{}
 The basis $\{\tilde{e}_\alpha\}$ will be used in the place of the $\{\ket{\alpha}\}$.
\[
{\bf{I}} \equiv \sum_{\alpha=1}^{d_0'} \bigl\|\,\mathcal{K}\lvert \alpha\rangle\,\bigr\|_2^{\,2}
   \;=\; \sum_{\alpha=1}^{d_0} \bigl\| [K,\;\tilde e_\alpha]\bigr\|_2^{\,2}
   \;=\; \frac{d_0}{d_0'} \sum_{\alpha=1}^{d_0}
        \operatorname{Tr}\!\left\{(K e_\alpha-e_\alpha K)\,(e_\alpha^\dagger K-K e_\alpha^\dagger)\right\}
\]
\[
=\; \frac{d_0}{d_0'} \sum_{\alpha=1}^{d_0}
   \operatorname{Tr}\!\left\{K e_\alpha e_\alpha^\dagger K
                           -K e_\alpha K e_\alpha^\dagger
                           -e_\alpha K e_\alpha^\dagger K
                           +e_\alpha K^2 e_\alpha^\dagger \right\}
\]
\[
=\; \frac{2d_0}{d_0'}\!\left(\,\|K\|_2^{\,2}-\langle K,\,P_{A'}(K)\rangle\right)
\;=\; \frac{2d_0}{d_0'}\!\left(\,\|K\|_2^{\,2}-\|P_{A'}(K)\|_2^{\,2}\right).
\]

\[
{\bf{II}}=\sum_{\alpha=1}^{d_0}\sum_{\beta=1}^{d_0'}\bigl|\langle\alpha|\mathcal{K}|\beta\rangle\bigr|^{2}
 =-\sum_{\alpha=1}^{d_0}\sum_{\beta=1}^{d_0'}
   \operatorname{Tr}\!\bigl(\tilde e_\alpha^\dagger\,\mathcal{K}(\tilde e_\beta)\bigr)
   \operatorname{Tr}\!\bigl(\mathcal{K}(\tilde e_\beta^\dagger)\,\tilde e_\alpha\bigr)
=-\sum_{\alpha=1}^{d_0}\sum_{\beta=1}^{d_0'}
  \operatorname{Tr}\!\Bigl[
   \bigl(\tilde e_\alpha\otimes \tilde e_\alpha^\dagger\bigr)\,
   \mathcal{K}^{\otimes 2}\!\bigl(\tilde e_\beta\otimes \tilde e_\beta^\dagger\bigr)
  \Bigr]
  \]
  \[=- \operatorname{Tr}\bigl(\tilde{\Omega}_A {\mathcal K}^{\otimes\,2}  \tilde{\Omega}_A   \bigr)=- \operatorname{Tr}\bigl(S\,{\Omega}_{A'} {\mathcal K}^{\otimes\,2}  \tilde{\Omega}_A   \bigr)=
  -\frac{d_0}{d_0'} \operatorname{Tr}\bigl(S\,{\Omega}_{A'} {\mathcal K}^{\otimes\,2}  {\Omega}_A   \bigr)
  \]
  \[
=-\frac{d_0}{d_0'}\sum_{\alpha=1}^{d_0}\sum_{\beta=1}^{d_0'}
  \operatorname{Tr}\!\Bigl[
   S\,(f_\alpha\otimes f_\alpha^\dagger)\,
   \mathcal{K}\bigl(e_\beta)\otimes \mathcal{K}(e_\beta^\dagger)\bigr)
  \Bigr],
\]
where we use $S$ to move from the $\{\tilde{e}_\alpha\}$ basis to the $\{f_\alpha\}$,  and  $\mathcal{K}(X)^\dagger=- \mathcal{K}(X^\dagger).$
Now using Eqs.~(\ref{eq:properties}), commutativity of the $e$'s with the $f$'s, and the cyclic property of the trace:
\[
{\bf{II}}=-\;\frac{d_0}{d_0'}\sum_{\alpha=1}^{d_0}\sum_{\beta=1}^{d_0'}
  \operatorname{Tr}\!\bigl[f_\alpha\,\mathcal{K}(e_\beta)\,f_\alpha^\dagger\,\mathcal{K}(e_\beta^\dagger)\bigr]
=\;-\frac{d_0}{d_0'}\sum_{\alpha=1}^{d_0}\sum_{\beta=1}^{d_0'}
  \operatorname{Tr}\!\Bigl[
    f_\alpha\,(K e_\beta-e_\beta K)\,f_\alpha^\dagger\,
    (K e_\beta^\dagger-e_\beta^\dagger K)
  \Bigr]
\]
\[
=\;-\frac{d_0}{d_0'}\sum_{\alpha=1}^{d_0}\sum_{\beta=1}^{d_0'}
  \operatorname{Tr}\!\Bigl[
    f_\alpha K e_\beta f_\alpha^\dagger K e_\beta^\dagger
   -f_\alpha K e_\beta f_\alpha^\dagger e_\beta^\dagger K
   -f_\alpha e_\beta K f_\alpha^\dagger K e_\beta^\dagger
   +f_\alpha e_\beta K f_\alpha^\dagger e_\beta^\dagger K
  \Bigr]
\]
\[
=\;\frac{d_0}{d_0'}\sum_{\alpha=1}^{d_0}\sum_{\beta=1}^{d_0'}
  \operatorname{Tr}\!\Bigl[
     f_\alpha K f_\alpha^\dagger e_\beta\,e_\beta^\dagger K
    + f_\alpha K f_\alpha^\dagger K  e_\beta^\dagger  e_\beta
    -e_\beta^\dagger K e_\beta\,f_\alpha^\dagger K f_\alpha
    -e_\beta K e_\beta^\dagger\,f_\alpha^\dagger K f_\alpha
  \Bigr]
\]
\[
=\;\frac{2d_0}{d_0'}\Bigl(\,\langle P_A(K),K\rangle-\langle P_A(K),P_{A'}(K)\rangle\Bigr).
\]
{Now we bring the two terms together:}
\[
\frac{1}{2}\|dP_\theta\|_{\mathrm{HS}}^2={\bf{I}}-{\bf{II}} \;=\; \frac{2d_0}{d_0'}\Bigl(
 \|K\|_2^{\,2}-\|P_A(K)\|_2^{\,2}-\|P_{A'}(K)\|_2^{\,2}
 +\langle P_A(K),P_{A'}(K)\rangle
\Bigr)
\]
\[
=\;\frac{2d_0}{d_0'}\Bigl(
 \|K\|_2^2-\|(P_A+P_{A'}-P_A P_{A'})(K)\|_2^{\,2}\Bigr)
\;=\;\frac{2d_0}{d_0'}\,
 \Bigl\|\,(1-P_{A+A'})(K)\,\Bigr\|_2^{\,2}.
\]
We note that if $A+A'=\{a+a'\,/\, a\in A,\, a'\in A'\}$, then the subspace projection $P_{A+A'}$ is equal to $P_A+P_{A'}-P_A\,P_{A'}.$ 
Finally, 
\[
ds_\Theta^2 =\frac{1}{d^2}\, \|dP_\theta\|_{\mathrm{HS}}^2= \kappa^2\,\|Q(K)\|_2^2,\quad \kappa:= \frac{2}{d_0'},
\]
 where $Q:=1-P_{A+A'}$ and we have used $d^2=d_0 d_0'.$

\end{document}